\documentclass[
 aps,
 pra,
 amsmath,amssymb,
 showpacs,
 showkeys,
 reprint,%
twocolumn,
]{revtex4-2}

\usepackage{graphicx}
\usepackage{dcolumn}
\usepackage{bm}
\usepackage{physics}
\usepackage{dsfont}
\usepackage{bm}
\usepackage[caption=false]{subfig}

\usepackage[titletoc]{appendix}

\makeatletter

\begin{document}


\title[Implementing perceptron models with qubits]{Implementing perceptron models with qubits}
\author{R.C. Wiersema}%
\email{roeland.wiersema@student.ru.nl}
\author{H.J. Kappen}
\affiliation{Department  of Biophysics, Donders Institute, Radboud University,  The Netherlands}%

\date{\today}

\begin{abstract}
    \noindent \textbf{We propose a method for learning a quantum probabilistic model of a perceptron. By considering a cross entropy between two density matrices we can learn a model that takes noisy output labels into account while learning. Although some work has been done that aims to utilize the curious properties of quantum systems to build a quantum perceptron, these proposals rely on the \textit{ad hoc} introduction of a classical cost function for the optimization procedure. We demonstrate the usage of a quantum probabilistic model by considering a quantum equivalent of the classical log-likelihood, which allows for both a quantum model and training procedure. We show that this allows us to better capture noisyness in data compared to a classical perceptron. By considering entangled qubits we can learn nonlinear separation boundaries, such as \textsc{XOR}.}
\end{abstract}



\maketitle

\section{Introduction}

One of the goals of quantum machine learning is to integrate quantum physics with machine learning to develop novel algorithms for learning classical data, so-called quantum-inspired models \cite{Yang2004, Stoudemire2016, Biamonte2017, Xie2017,  Amin2018}. 
Along with these developments another goal has been to come up with machine learning algorithms for quantum computers, either by designing specific algorithms for quantum computers \cite{Neven2018, Schuld2019}, or by speeding up the underlying linear algebra routines \cite{Lloyd2009}. Examples of the former include employing adiabatic quantum annealers to train a binary classifier \cite{Neven2018} and using a quantum computer to calculate an classically intractable kernel function \cite{Schuld2019}, whereas the latter includes support vector machines \cite{Lloyd2014}, support matrix machines \cite{Bojia2017}, A-optimal projections \cite{Bojia2019}, and principal component analysis (PCA) \cite{Lloyd2013}. However, most of these proposals remain unfeasible due to the current limitations of modern quantum computers, which still lack long qubit (the quantum mechanical description of a single spin-$\frac{1}{2}$ particle) coherence times and high gate fidelity \cite{Preskill2018}. 

Inspired by the success of deep learning \cite{LeCun2015}, there has been interest to develop quantum equivalents of neural networks that can be trained more efficiently or are more expressive than their classical counterparts \cite{Schuld2014, Jeswal2018, Zhou1999, Kouda2005, Zhou2007, Shang2015, Schuld2015, Wan2017}. Quantum-inspired proposals utilize quantum effects in different ways: employing a superposition of perceptrons \cite{Zhou1999}, using qubit weights \cite{Kouda2005, Shang2015}, or learning a unitary transformation between input and output \cite{Zhou2007}. Quantum computing work in this direction involves using an inverse quantum Fourier transform to obtain a nonlinear step function \cite{Schuld2015} or tracing out parts of a quantum circuit to create an autoencoder \cite{Wan2017}. However, all these proposals introduce a classical cost function for learning, omitting the underlying probabilistic motivation for their model. The usage of quantum probabilistic cost functions is still relatively unexplored. \newline

Constructing quantum probabilistic models from density matrices is a recent direction of quantum machine learning research \cite{Amin2018, Kappen2018}, where one exploits quantum effects in both the model and training procedure by constructing a differentiable cost function in terms of density matrices. Density matrices are used in quantum mechanics to describe statistical ensembles of quantum states. They are represented by a positive semidefinite Hermitian matrix with trace 1. In this Rapid Communication, we will use density matrices to construct a model that minimizes a generalization of the classical likelihood function for learning, replacing the classical perceptron bit with a qubit. Others have attempted to generalize probability theory to density matrices \cite{Warmuth2006}. However, the equivalent of conditional probabilities, conditional density matrices, do not preserve positive definiteness so states can be assigned a negative probability \cite{Cerf1999}. Our approach bypasses this difficulty because we construct a data density matrix from the probability amplitude of the empirical data distribution, which is always positive semidefinite. 

The desired perceptron is a linear classifier that can be used for binary classification. It assigns a probability
\begin{equation}
    p(y=1|\mathbf{x}) = f(\mathbf{x}\cdot \mathbf{w})
\end{equation}
to class $y=1$, based on input $\mathbf{x}$ and trainable weights $\mathbf{w}$ with $f(x)$ a nonlinear activation function. The activation function of the perceptron is often taken to be a sigmoid, since it produces an output between $0$ and $1$ and is equivalent to logistic regression. The perceptron is of particular interest in machine learning because it is the building block of multilayer neural networks, the driving force behind deep learning.

In Sec. \ref{sec:qperceptron} we will consider a qubit perceptron that uses a generalization of the classical likelihood function for learning. Some numerical results for toy data sets are discussed in Sec. \ref{sec:results}, where we show that our qubit model is better at assigning class probability for noisy data. In Sec. \ref{sec:entangled_perceptron} we will consider two entangled qubits as a perceptron that can learn nonlinear problems by assigning a nonlinear separation boundary.

\section{Quantum Perceptron}\label{sec:qperceptron}
Consider a classification problem where we have a data set consisting of input vectors $\mathbf{x}\in \mathbb{R}^d$ of length $d$ with corresponding labels $y\in \{1,-1\}$. In supervised machine learning it is our goal is to find the parameters $\mathbf{w}$ for the function $p(y|\mathbf{x}; \mathbf{w})$ that assigns a high probability to the correct label $y$ for each input $\mathbf{x}$. The classical negative log-likelihood is given by
\begin{equation}
   \mathcal{L}_{cl} = -\sum_{\mathbf{x}} q(\mathbf{x}) \sum_y q(y|\mathbf{x})\ln p(y|\mathbf{x}; \mathbf{w}).
\label{eq:cl_lh}
\end{equation}
Here, $q(\mathbf{x})$ is the empirical probability of observing $\mathbf{x}$, $q(y|\mathbf{x})$ is the empirical conditional probability of observing label $y$ for data $\mathbf{x}$, and $p(y|\mathbf{x}, \mathbf{w})$ is the proposed model conditional probability distribution of the data. By performing a gradient descent we can find the optimal parameters for our model, which is equivalent to minimizing the cross entropy between distributions $p$ and $q$.

To extend the classical likelihood in Eq. (\ref{eq:cl_lh}) to the realm of quantum mechanics we require a description of our model and the conditional probability $q(y|\mathbf{x})$ in terms of density matrices. The density matrix contains the classical uncertainty we have about a quantum state. If this matrix is rank one, we have what is known as a pure state in which case there is no classical uncertainty about what quantum state the system is in. If the density matrix has rank $>1$, then we have a so-called mixed state \cite{Nielsen2011}. For our model we will consider a parametrized mixed state, since this will allow us to capture the uncertainty in the data. To perform learning, we require a learning rule that preserves the Hermiticity, positive semidefiniteness and trace of the density matrix.

We consider the specific case where the data consist of $N$ discrete vectors $\mathbf{x}\in \{1, -1\}^d$ with $d$ bits and $y \in \{1, -1\}$ labels. We define the quantum log-likelihood as a cross entropy between a conditional data density matrix $\eta_\mathbf{x}$ and a model conditional density matrix $\rho_\mathbf{x}$, analogous to Eq. (\ref{eq:cl_lh}). For each $\mathbf{x}$ we construct a wave function based on the empirical conditional probabilities $q(y|\mathbf{x})$,
\begin{equation}
    \ket{\Psi} = \sqrt{q(1|\mathbf{x})}\ket{1} + \sqrt{q(-1|\mathbf{x})}\ket{-1},
\end{equation}
where the states $\ket{1}$, $\ket{-1}$ are the eigenstates of the $\sigma^z$ operator. The data density matrix is defined as $\eta_\mathbf{x} \equiv \ketbra{\Psi}{\Psi}$, with components
\begin{equation}
    \eta_\mathbf{x}(y, y^\prime) = \sqrt{q(y|\mathbf{x})}\sqrt{q(y^{\prime}|\mathbf{x})}\label{eq:data_densit}.
\end{equation}
Note that this is a pure density matrix. $q(y|\mathbf{x})$ is an empirical distribution over the label $y$ for each $\mathbf{x}$, and is fully determined by its conditional expectation value of $y$ given $\mathbf{x}$ written as $b(\mathbf{x})$,
\begin{align}
    q(y|\mathbf{x}) &= \frac{1}{2}(1+b(\mathbf{x}) y)\label{eq:sample_prob},\\
    \text{with } b(\mathbf{x})&= \frac{1}{M}\left(\sum_{\mathbf{x}^\prime} 
    y^{\prime}\mathds{I}(\mathbf{x}^\prime=\mathbf{x})\right)\nonumber\\
    \text{and } M &= \sum_{\mathbf{x}^\prime}\mathds{I}(\mathbf{x}^\prime = \mathbf{x}).\nonumber
\end{align}
Succinctly put, every time $\mathbf{x}$ appears in the data, we add its corresponding label $y^\prime$ to the sum. Dividing by $M$, the total number of times the sample $\mathbf{x}$ appears in the data we obtain the conditional expectation value $b(\mathbf{x})$. We define the empirical probability 
\begin{align*}
    q(\mathbf{x}) = \frac{M}{N}\\
\end{align*}
for $M$ occurrences of $\mathbf{x}$ and $N$ the total number of samples. 

Our model is a density matrix $\rho(\mathbf{x}, \mathbf{w}; y, y^\prime)\equiv \rho_\mathbf{x}$. We use the following proposal,
\begin{equation}
    \rho_\mathbf{x} = \frac{1}{Z} e^{-\beta H}\label{eq:density},
\end{equation}
where $H = \sum_k h^k \sigma^k$, with $h^k \in \mathbb{R}$ and $\sigma^k$ the Pauli matrices with $k = (x,y,z)$. This is a finite temperature description of a qubit, where we will set $\beta=-1$ for now. Using that $\exp({a\:\hat{\mathbf{n}}\cdot \bm{\sigma}})= \cosh(a) + \sinh{(a)}\sum_k \sigma^k $ and writing $\sum_k h^k \sigma^k = h \sum_k \frac{h^k}{h} \sigma^k = $ with $h = \sqrt{\sum_k(h^k)^2}$, we find
\begin{equation}
    \rho_\mathbf{x} = \frac{1}{Z}\left( \cosh h + \sinh h\sum_k \frac{h^k \sigma^k}{h}\right).
\end{equation}
Solving $\Tr{\rho_\mathbf{x}}=1$ gives $Z = 2 \cosh h$. Then,
\begin{align}
    \rho_\mathbf{x} & = \frac{1}{2} I + \frac{1}{2} \tanh h\sum_k \frac{h^k \sigma^k}{h}\nonumber \\
    &= \frac{1}{2} I + \frac{1}{2} \sum_k m^k \sigma^k, \label{eq:density_matrix_basis}
\end{align}
where $I$ is a $2\times2$ identity matrix and $m^k = \frac{h^k}{h}\tanh h$. Eq. (\ref{eq:density_matrix_basis}) gives us the general description of a qubit, which we have now described in terms of a density matrix. This definition spans the space of $2\times2$ Hermitian matrices, for all $h^k\in \mathbb{R}$. From the definition of $m^k$ it is clear that $m^k\in(-1,1)$. This means that $\rho_\mathbf{x}$ is positive semidefinite because the eigenvalues of $\rho_\mathbf{x}$ are 
\begin{equation}
   \lambda_{\pm} = \frac{1}{2}(1 \pm \sqrt{ \sum_k (m^k)^2}) \geq 0.
\end{equation}
From the eigenvalues we also see that $\rho_\mathbf{x}$ describes a mixed state, since it is only rank one if $\sum_k (m^k)^2 = 1$. \newline

We now parametrize the field $h^k\to h^k(\mathbf{x})$ by setting $h^k(\mathbf{x}) = \mathbf{w}^k\cdot \mathbf{x}$ with $\mathbf{w}^k \in \mathbb{R}^d$, so that the qubit state is dependent on classical input data. We can absorb the inverse temperature $-\beta$ in the field $-\beta h^k \to h^k$ by rescaling the weights $\mathbf{w}^k$. Note that for each Pauli matrix $k$, we have one set of weights $\mathbf{w}^k$. To clean up the notation we omit the argument of $h^k$ from now on. We now generalize Eq. (\ref{eq:cl_lh}) with our data and model density matrices $\eta_\mathbf{x}$ and $\rho_\mathbf{x}$ to obtain the negative quantum log-likelihood,
\begin{equation}
    \mathcal{L}_{q} = -\sum_{\mathbf{x}} q(\mathbf{x}) \Tr{\eta_\mathbf{x}\ln(\rho_\mathbf{x})}.
    \label{eq:q_like}
\end{equation}
This is the quantum mechanical equivalent of the classical log-likelihood which minimizes the ``distance" between the density matrix representations of the data and the model. This expression also appears in the quantum relative entropy, and for $\eta_\mathbf{x}>0$ the quantum log-likelihood is convex in $\rho_\mathbf{x}$ \cite{Carlen2010}. Next, we rewrite this with our parametrized $\rho_\mathbf{x}$,
\begin{align}
     \mathcal{L}_{q} =& -\sum_{\mathbf{x}} q(\mathbf{x}) \Tr{\eta_\mathbf{x}\ln(\rho_\mathbf{x})} \\
    =& -\sum_{\mathbf{x}} q(\mathbf{x}) \sum_{y,y^\prime} \bra{y\prime}\sqrt{q(y|\mathbf{x})} \nonumber \sqrt{q(y^{\prime}|\mathbf{x})}\ln(\rho_\mathbf{x})\ket{y}
\end{align}
with $\{\ket{y}\}$ a set of orthonormal vectors in the $\sigma^z$ basis,
\begin{align}
    -\sum_{\mathbf{x}} & q(\mathbf{x}) \sum_{y,y^\prime} \sqrt{q(y|\mathbf{x})} \sqrt{q(y^{\prime}|\mathbf{x})} \nonumber\\ 
    & \times\bra{y^\prime}\left(\sum_k h^k \sigma^k - \ln(2\cosh h)\right)\ket{y}.
    \label{eq:intermed_step}
\end{align}
Calculating the statistics for the Pauli matrices gives
\begin{align}
    \sum_{y,y^\prime}\bra{y^\prime} \sum_k h^k \sigma^k \ket{y} &= \sum_{y, y^\prime}\sum_k \bra{y^\prime}h^k \sigma^k \ket{y},
\end{align}
which gives three delta functions that we can plug into Eq. (\ref{eq:intermed_step}) together with our definition of $q(y|\mathbf{x})$ from Eq. (\ref{eq:sample_prob}),
\begin{align}
    &\sum_{y, y^\prime} \sqrt{q(y|\mathbf{x})} \sqrt{q(y^{\prime}|\mathbf{x})} \left( h^x \delta_{y^\prime,-y} + i y h^y \delta_{y^\prime,-y} + y h^z \delta_{y^\prime,y} \right)\nonumber\\
    &=h^{x} \sqrt{1 - b(\mathbf{x})^2} + h^z b(\mathbf{x}).
\end{align}
The $h^x$ term quantifies how often a sample occurs with a flipped output label and is the distinguishing factor from the classical perceptron. The source of this term is the $\sigma^x$ matrix in the likelihood which flips the state $\ket{y}$ and scales $h^x$ with the off-diagonal elements of $\eta_\mathbf{x}$. As a final likelihood we get
\begin{align}
    \mathcal{L}_{q} = -\sum_{\mathbf{x}} q(\mathbf{x})  &\bigg(h^x \sqrt{1 - b(\mathbf{x})^2} + h^z b(\mathbf{x}) \nonumber\\
    &- \ln(2\cosh h)\bigg)\label{eq:final_likelihood}.
\end{align}
In order to perform learning we have to find update rules that minimize the function in Eq. (\ref{eq:final_likelihood}). To find the minimum we perform a gradient descent to update the parameters $\mathbf{w}^k$. Derive with respect to $\mathbf{w}^k$,
\begin{align}
    \frac{\partial \mathcal{L}_{q}}{\partial \mathbf{w}^x} &= -\sum_{\mathbf{x}} q(\mathbf{x})  \left(\sqrt{1 - b(\mathbf{x})^2}  - \frac{h^x}{h} \tanh h \right)\mathbf{x},\nonumber\\
    \frac{\partial \mathcal{L}_{q}}{\partial \mathbf{w}^y} &= \sum_{\mathbf{x}} q(\mathbf{x})  \left(\frac{h^y}{h} \tanh h\right)  \mathbf{x},\nonumber\\
    \frac{\partial \mathcal{L}_{q}}{\partial \mathbf{w}^z} &= -\sum_{\mathbf{x}} q(\mathbf{x})  \left( b(\mathbf{x}) - \frac{h^z}{h} \tanh h \right)\mathbf{x}.
\end{align}
Update the weights at iteration $t$ with
\begin{align}
    \mathbf{w}^k(t+1) =  \mathbf{w}^k(t) - \epsilon \left(\frac{\partial \mathcal{L}}{\partial \mathbf{w}^k(t)}\right).
    \label{eq:grad_desc}
\end{align}
These are the learning rules for the quantum perceptron, with learning parameter $\epsilon$ for each gradient. Since the gradient step of $\mathbf{w}^y$ is proportional to $\mathbf{w}^y$, the fixed-point solution is $\mathbf{w}^y\rightarrow\bm{0}$ in the limit of many iterations. In the case that there exists a function $f(\mathbf{x}) = y$ (no noise in the data) for all data points, the statistics $b(\mathbf{x})$ become either $1$ or $-1$, which gives a fixed-point solution $\mathbf{w}^x\rightarrow\bm{0}$. The $h^z$ field then corresponds to the single field of a classical perceptron and the quantum perceptron approaches the classical case. However, in the case where there are samples which have both $1$ and $-1$ labels, the weight $\mathbf{w}^x$ becomes finite and the solution of the quantum perceptron will diverge from the classical perceptron. This change in behavior is reflected in the probability boundaries, which differ from the classical case (see appendix \ref{sec:boundaries}). \newline

We have yet to address how we actually retrieve the a class label $y$ from the model. Once trained, we can construct a state $\rho_\mathbf{x}$ of the qubit based on some input $\mathbf{x}$. The output labels $y \in \{-1,1\}$ correspond to the states $\ket{-1}$, $\ket{1}$ by construction. An obvious measure of probability is the expectation value $\expval{\sigma^z}_{\rho_\mathbf{x}}$, which gives $p(y|\mathbf{x}; \mathbf{w}) = \frac{1}{2}(1 + y\expval{\sigma^z}_{\rho_\mathbf{x}})$. For a finite-temperature system we have for the expectation value of some observable $\Hat{A}$,
\begin{equation}
    \expval{\Hat{A}} = \Tr{\Hat{A} \rho}.
\end{equation}
From our definition in Eq. (\ref{eq:density_matrix_basis}) we see that
\begin{equation}
    \expval{\sigma^z}_{\rho_\mathbf{x}} = \Tr{\sigma^z\frac{1}{2}(1 + \sum_k m^k \sigma^k)} = \delta_{kz} m^k = m^z,
\end{equation}
where we used that $\Tr{\sigma^i}=0$ and $\Tr{\sigma^i\sigma^j}=2\delta_{ij}$. The class probability is then constructed as 
\begin{equation}
    p(y|\mathbf{x}; \mathbf{w}) = \frac{1}{2}(1 + y m^z) \label{eq:probability}.
\end{equation}

\section{Results}\label{sec:results}
In this section we apply the quantum perceptron to some toy data sets and compare with the classical perceptron with a sigmoid activation function, i.e., logistic regression. For both the classical and quantum perceptron we look at the mean squared error (MSE) to evaluate the performance of both methods,
\begin{equation}
    \text{MSE} = \frac{1}{N}\sum_i^N \left(y_i - p(y_i|\mathbf{x_i};\mathbf{w})\right)^2.
\end{equation}
We always reach the global minimum through batch gradient descent because the cost functions are convex for both models. Due to the flatness of the likelihood function near the global minimum, convergence can be slow. Setting the threshold for convergence at $\Delta \mathcal{L} <10^{-7}$ and the learning parameter at $\epsilon=0.01$ ensures that we obtain fast convergence without sacrificing model accuracy for the problems discussed in this Rapdid Communication.

\subsection {Two-dimensional binary problem}

In order to demonstrate the difference between the classical and quantum perceptron we consider a two-dimensional binary classification problem. If the problem is linearly separable, the classical perceptron converges to a solution where the two classes are perfectly separated. In the case where some samples are `mislabeled' the quantum perceptron should behave differently, because we account for noise in the learning rule.

\begin{figure}[htb!]
    \subfloat{%
        \includegraphics[clip,width=0.94\columnwidth]{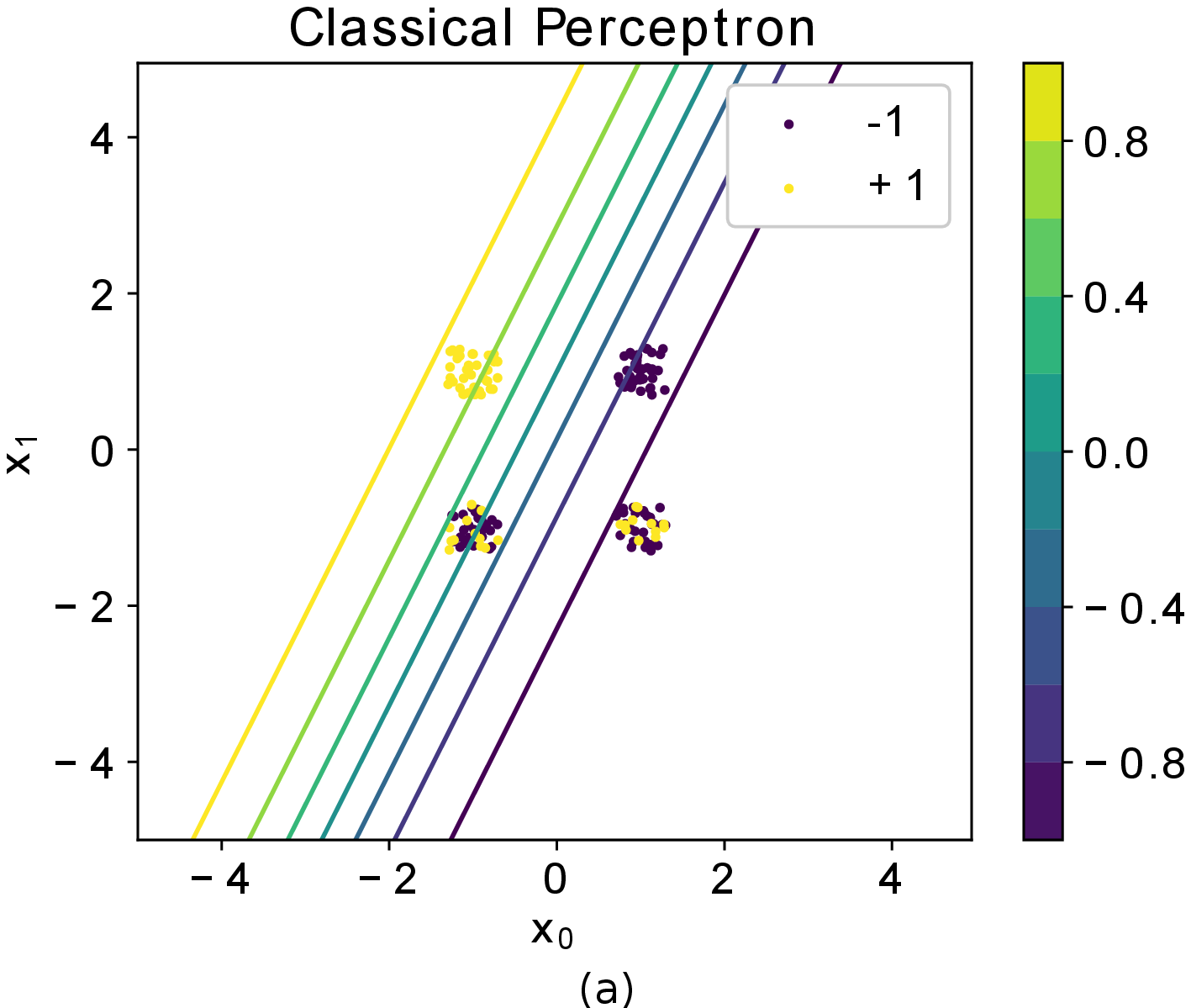}
    }
    
    \subfloat{%
        \includegraphics[clip,width=0.94\columnwidth]{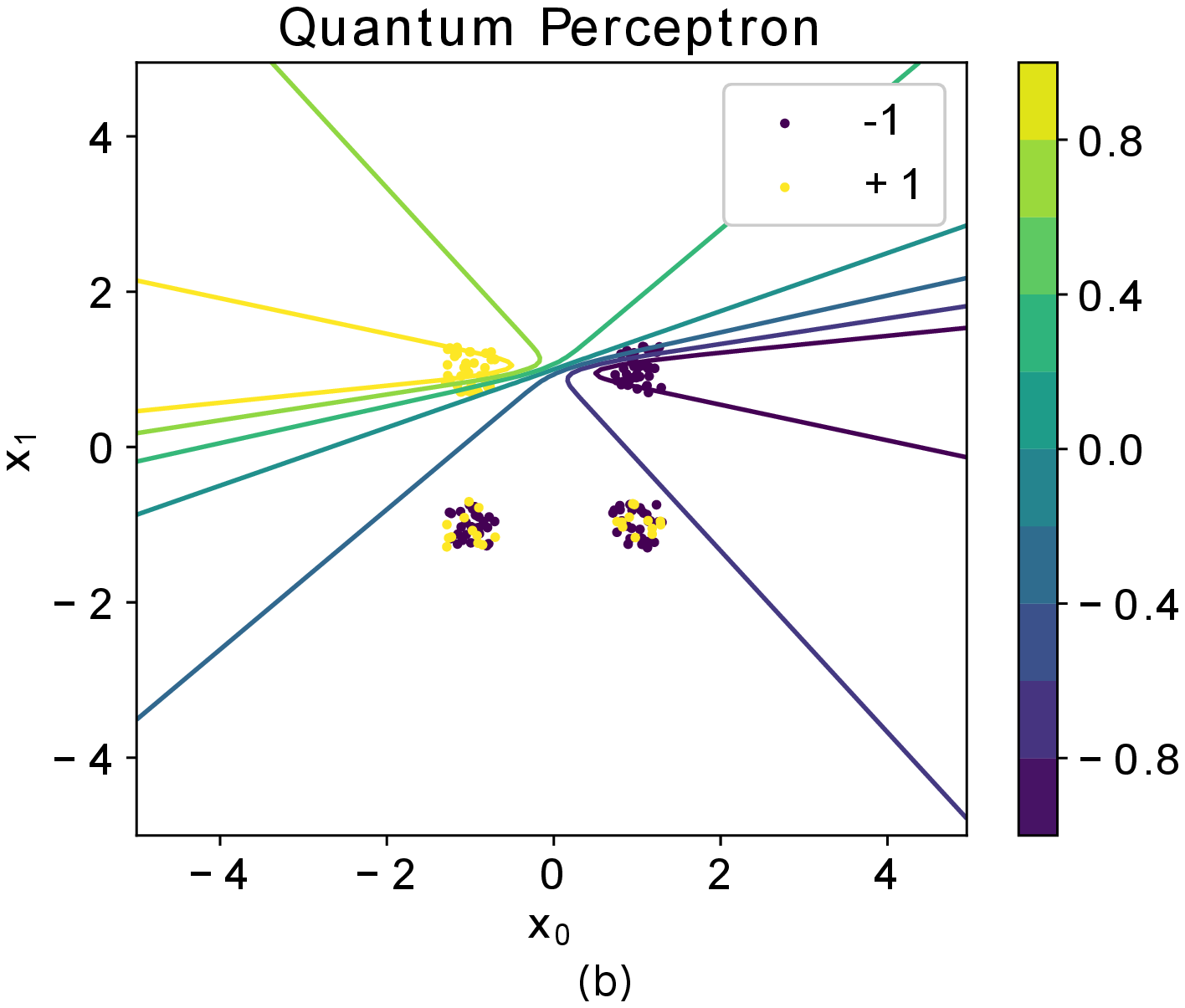}
    }
    \caption{Separation boundaries in the input space for a two dimensional problem with $\mathbf{x}=(x_0, x_1)$. The contour lines indicate the expectation value $\mathbb{E}[y|\mathbf{x};\mathbf{w}]\in(-1,1)$. The $0.0$ line indicates the separation boundary where $p(y=1|\mathbf{x};\mathbf{w}) = p(y=-1|\mathbf{x};\mathbf{w})=\frac{1}{2}$. Random jitter is added to the plot to clarify which samples are noisy. \textbf{(a)} Classical perceptron. The classical perceptron assigns linear boundaries through the input space, where the distance between the boundaries is scaled with the sigmoid. \textbf{(b)} Quantum perceptron. The quantum perceptron assigns curved boundaries through the input space. Samples with mislabelings get assigned a lower expectation value which results in a lower MSE of $\text{MSE}(quantum) \approx 0.106$ for the quantum perceptron vs $\text{MSE}(classical) \approx 0.154$ for the classical perceptron. Note that if we threshold the quantum perceptron boundary at $p(y=1|\mathbf{x}; \bm{\theta}) = 0.5$, we get a linear boundary that would assign similar classes as in (a), even though the boundary is tilted with respect to the classical boundary. However, the quantum perceptron assigns high probabilities to classes about which it is certain ($\mathbf{x}\in \{(-1,1),(1,1)\}$) and lower probabilities to classes about which it is uncertain ($\mathbf{x}\in \{(-1,-1),(1,-1)\}$). The classical perceptron does this significantly worse, which is reflected in the difference in MSE.}
    \label{fig:boundaries_qm}
\end{figure}

Consider the data $\mathbf{x} = \{(1,1), $ $(1,-1), $ $(-1,1), $ $(-1,-1) \}$  with labels $y = \{-1,-1,1,-1\}$ respectively. This problem is trivial since it is linearly separable and all algorithms converge to the same solution ($\mathbf{w}^{x,y}=\mathbf{0}$ and $\mathbf{w}^z \approx \mathbf{w}_{cl}$). However, if we flip some of the output labels to simulate mislabeled samples or errors in the data, we suspect that the quantum perceptron will perform better. We make $40$ copies of the four data points in the binary feature space and for $\mathbf{x}\in \{(1,-1),(-1,-1)\}$ we flip $30\%$ of the outputs from $-1$ to $1$. The probability boundaries of the perceptrons differ significantly, as can be seen in Fig. \ref{fig:boundaries_qm}, which leads to a better assignment of the probability of the correct states. An explanation for the shape of the boundaries can be found in appendix \ref{sec:boundaries}.
\captionsetup[subfigure]{labelformat=empty}

\subsection{Binary teacher-student problem}

\begin{figure}[htb!]
    \centering
    \includegraphics[width=1.1\linewidth]{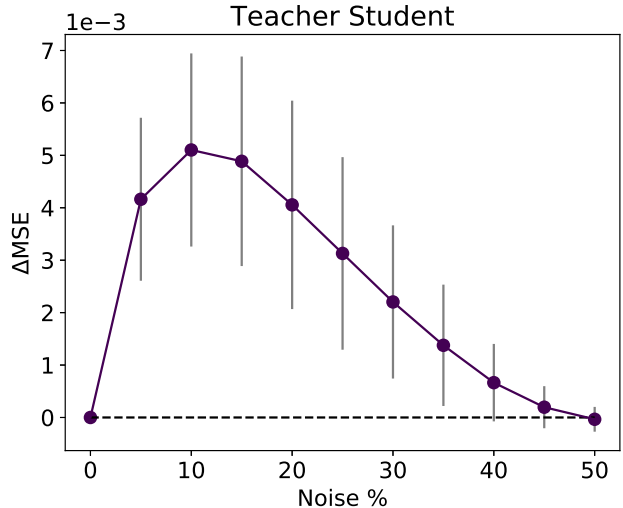}
    \caption{$\Delta\text{MSE} = \text{MSE}(classical) - \text{MSE}(quantum)$ vs the percentage of labels flipped in the training data. Error bars indicate the standard deviation over 100 different $\mathbf{w}_{teacher}$ initializations. If the amount of noise is $0\%$, the classical and quantum perceptron will converge to the same solution. If the amount of noise is $50\%$, then both models cannot learn anything. Between these two points lies an area where the quantum perceptron outperforms the classical perceptron.} 
    \label{fig:lh_diff_mz}
\end{figure}

A more complex, higher-dimensional problem is the teacher-student problem. The input data $\mathbf{x}\in \mathbb{R}^d$ consist of 600 random binary vectors of length $d=8$, where $\mathbf{x}\in\{-1, 1\}^d$. We take a random weight vector $\mathbf{w}_{teacher} \sim \mathcal{N}(0,1)$ and determine labels $y = \text{sgn}(\mathbf{x} \cdot \mathbf{w}_{teacher})$. We then create five duplicates of each input vector to ensure that there are multiple copies of each sample in the data set. Next, we flip some percentage of labels for $80\%$ of this data set (the training set). This is done by generating a random permutation of the indices of the samples and flipping the label for the first $x\%$ of them. After training both the classical and quantum perceptron we predict the labels for the remaining $20\%$ of the data (the test set) and calculate the difference in MSE between the two models. The percentage of flipped labels was incrementally increased by $5\%$ from $0\%$ to $50\%$. At each step in this schedule we learn 100 different $\mathbf{x}$ and $\mathbf{w}_{teacher}$ to gather statistics for the mean and variance of $\Delta$MSE. The 100 generated problems are equal across the different percentages.
This setup allows us to assert whether the algorithms can still find the original separation of the data even if noise is introduced. The performance of the quantum perceptron and classical perceptron is compared in Fig. \ref{fig:lh_diff_mz}.\newline

We have shown that the quantum probabilistic description is better than a classical perceptron in capturing uncertainty in toy data sets. At the cost of introducing an additional parameter $\mathbf{w}^x$, the model is more expressive, which allows for a better characterization of the data.

\section{Entangled Perceptron}\label{sec:entangled_perceptron}
\begin{figure}[ht]
    \centering
    \includegraphics[width=1.1\linewidth]{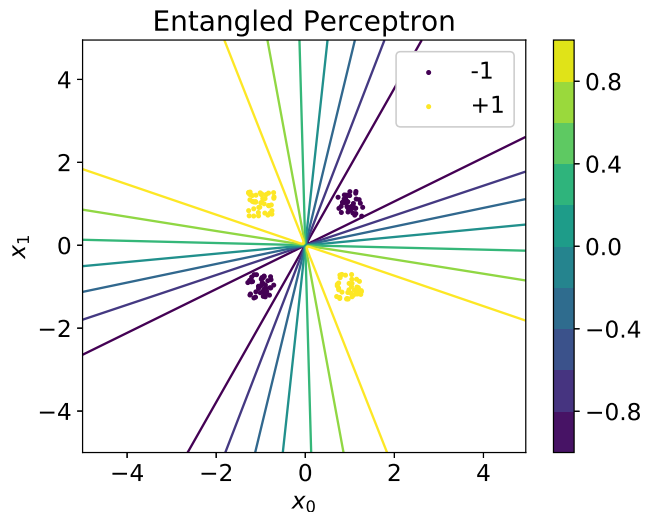}
    \caption{The \textsc{XOR} problem. Perfect classification of this nonlinear data set requires four classical perceptrons in a two-layer configuration or a kernel transformation $(x_0,x_1)\to(x_0,x_1,\sqrt{x_0^2+x_1^2})$. We show that the problem can be learned perfectly with two qubits.} 
    \label{fig:entan_percep}
\end{figure}

In this section we demonstrate the use of entanglement for learning. This can be achieved by extending the previous ideas to a multiqubit system. Consider the Hilbert space $\mathcal{H} = \mathcal{H}_A \otimes\mathcal{H}_B$, with $i,j=0,1$. Let $\{\ket{\phi_i}\}$ be an orthonormal basis for the $2\times 2$ Hilbert spaces $\mathcal{H}_{A}$ and $\mathcal{H}_{B}$. We can write down an arbitrary state in $\mathcal{H}$ as
\begin{equation}
    \ket{\phi} = \frac{1}{\sqrt{N}}\sum_{i,j} h^{ij} \ket{\phi_i}\otimes\ket{\phi_j},
\end{equation}
where $h^{ij}\in \mathbb{C}$. We must normalize $\ket{\phi}$ accordingly to ensure that $\braket{\phi}{\phi} = 1$, with $\braket{\phi}{\phi} = \sum_{ij} h^{ij*}h^{ij} \equiv N$. This state can be described with a density matrix that is rank one because we are dealing with a pure state. Since $\rho \neq \rho_A\otimes\rho_B$ in general the state can be entangled. If we now look at the reduced density matrix $\rho_B$ by tracing out qubit $A$ we end up with a mixed state.
\begin{align}
    \rho_B &=\frac{1}{N} \sum_{i,j, j^\prime} h^{ij*}h^{i j^\prime}\ket{\phi_j}\bra{\phi_{j^\prime}} \label{eq:rho_red}
\end{align}
If we take $h^{ij} = \mathbf{w}^{ij}\cdot \mathbf{x}$ with $\mathbf{w}^{ij}\in \mathbb{C}^d$, then we have constructed a quantum state parametrized by our inputs. With the data density matrix we used in Eq. (\ref{eq:data_densit}) we can again minimize the quantum log-likelihood in Eq. (\ref{eq:q_like}) by replacing $\rho_\mathbf{x}$ with $\rho_B$. We can now learn nonlinear problems as can be seen in Fig. \ref{fig:entan_percep}. An explanation of the quadrics and the shape of the boundaries as well as additional examples can be found in appendix \ref{sec:entangled_q}.

\section{Conclusion}\label{sec:conclusion}

We extended the classical likelihood to a quantum log-likelihood and constructed a quantum perceptron from density matrices. The resulting algorithm is more resistant to noisy data when learning and takes this noisiness into account when predicting. This is due to the fact that there is a cost for flipped output labels in the quantum log-likelihood. For toy data sets we observed that the quantum perceptron is better at assigning probability to noisy samples, which resulted in improved performance. When we considered the extension to two entangled qubits, we could also learn nonlinear separation boundaries.

In this Rapid Communication we have only considered binary classification, but the quantum perceptron can easily be extended to multiclass regression for $C>2$ classes by considering the $\text{SU}(C)$ generators instead of the Pauli matrices. These generators span the space of $C\times C$ traceless Hermitian matrices. We are then working with qudits, which generalize the properties of qubits to $d$-level quantum systems.

A caveat of the quantum perceptron is that in order to outperform a classical perceptron, we require multiple copies of a sample $\mathbf{x}$ with conflicting labels $y$ to be present in the data, otherwise $b(\mathbf{x})= \pm 1$ for all data points and the algorithm simply reduces to the classical perceptron. Most real-world data sets, however, contain either a large number of features or continuous variables so that copies of samples $\mathbf{x}$ with conflicting labels are rare. This limits the increase in performance to edge cases where there is discrete input data with a small number of features. Nonetheless, we have shown that at the cost of introducing a single parameter $\mathbf{w}^x$, the density matrix construction is a more general model than the classical perceptron.

To conclude, we have shown that it is possible to learn a quantum model using a quantum cost function and that this can lead to improved performance for toy data sets. We believe that this modeling paradigm could be a fruitful direction for developing algorithms for noisy intermediate scale quantum computers, since the quantum probabilistic approach is still relatively unexplored in the current literature. The code with the \textsc{Tensorflow} model of the quantum perceptron and ways to reproduce the figures in this Rapid Communication can be found on GitHub \cite{Wiersema2019}.

\section{Acknowledgment}
We thank J. Mentink and the people involved with the ``Bits and Brains" project for inspiring discussions. This research was funded in part by ONR Grant N00014-17-1-256.

\bibliographystyle{unsrt}
\bibliography{library.bib}

\begin{appendices}
\section*{\label{appendix} Appendix}

\subsection{Single qubit probability boundaries}\label{sec:boundaries}

\noindent We can analyze the separation boundaries learned by our model. Setting
\begin{equation}
    p(y=1|\mathbf{x};\mathbf{w}) = p(y=-1|\mathbf{x};\mathbf{w})
\end{equation}
gives the boundaries of equal probability. Plugging in our definition of class probability
\begin{align}
    \frac{1}{2}(1 + m^z) &= \frac{1}{2}(1 - m^z)\nonumber\\
    m^z& = \frac{h^z}{h} \tanh h= 0\label{eq:hz_h}
\end{align}
which is solved for $h^z=0$, giving a $n-1$ dimensional hyperplane, just as for a classical perceptron. We can also analyze curves of equal probability in the input space. Algebraically this corresponds to
\begin{equation}
    p(y=1|\mathbf{x};\mathbf{w}) = p(y=-1|\mathbf{x};\mathbf{w})+ \epsilon  \label{eq:p_eps}
\end{equation}
with $\epsilon\in [0,1]$. Using that $\sum_y p(y|\mathbf{x};\mathbf{w}) = 1$ we get
\begin{align}
    p(y=1|\mathbf{x};\mathbf{w}) &= \frac{1}{2}(1 + \epsilon)\label{eq:curve}
\end{align}
In the limit that $h^y \to 0$ this gives
\begin{align}
    \frac{h^z}{\sqrt{(h^x)^2+(h^z)^2}} \tanh h &= \epsilon\nonumber\\
    (h^z)^2 \tanh^2 h&= ((h^x)^2+(h^z)^2) \epsilon^2 \label{eq:mz_epsilon}
\end{align}
For large $h$ we have $\tanh h\approx 1$.
\begin{align}
    h^z &= \pm\delta h^x
\end{align}
with $\delta = \epsilon^2 / \sqrt{1- \epsilon^2}$. This gives a hyperplane equation 
\begin{equation}
    \mathbf{w}^z \cdot \mathbf{x} \pm \delta \mathbf{w}^x \cdot \mathbf{x}= 0
\end{equation}
For $\delta\neq0$ we require that
\begin{align}
    \mathbf{w}^x \cdot \mathbf{x} + w_0^x &= 0 \nonumber\\
    \mathbf{w}^z \cdot \mathbf{x} + w_0^z &= 0
\end{align}
Both these equations do not depend on $\delta$, so these $\delta$-hyperplanes intersect in the same subspace. Assuming that $\mathbf{w}^x$ and $\mathbf{w}^z$ are linearly independent, we can solve for 2 of the $n$ variables in $\mathbf{x}$. The two $\pm \delta$ solutions intersect in a $n-2$ dimensional subspace and both span a $n-1$ dimensional hyperplane.

\subsection{Entangled qubit probability boundaries}\label{sec:entangled_q}

We can identify entries of the reduced density matrix with the ones from the single qubit and reuse the analysis done in the previous section.
\begin{align*}
    \frac{1}{2}(1+m^z) &= \frac{h_{00}^2 + h_{10}^2}{N}\\
    m^z &= \frac{2(h_{00}^2 + h_{10}^2)}{N} - 1
\end{align*}
Solving $m^z = 0$ analogous to Eq. (\ref{eq:hz_h})
\begin{align}
    &2(h_{00}^2 + h_{10}^2) = h_{00}^2 + h_{10}^2 + h_{01}^2 + h_{11}^2\nonumber\\
    &h_{00}^2 + h_{10}^2 - h_{01}^2 - h_{11}^2 = 0 \label{eq:hypersurface}
\end{align}
The square of a dot product can be written as
\begin{align}
    h_{ij} h_{kl} &= (\mathbf{w}_{ij} \cdot \mathbf{x}) (\mathbf{w}_{kl} \cdot \mathbf{x}) \nonumber\\
    & = \sum_{\mu, \nu} {w}_{ij}^\mu \mathbf{x}^\mu {w}_{kl}^\nu \mathbf{x}^\nu = \mathbf{x}^T A_{ijkl} \mathbf{x}
\end{align}
If $A$ is symmetric, then $\mathbf{x}^TA\mathbf{x}$ is a quadratic form. However, the form $ \mathbf{x}^TA_{ijkl}\mathbf{x}$, is not symmetric since in general ${w}_{ij}^\mu \neq {w}_{kl}^\nu$. We can redefine
\begin{align}
    {w}^{sym}_{ijkl} &= \frac{1}{2}({w}^{0}_{ij}{w}^{1}_{kl} + {w}^{1}_{ij}{w}^{0}_{kl})\\
    {w}^{00}_{ijkl} &= {w}_{ij}^0 {w}_{kl}^0\nonumber\\
    {w}^{11}_{ijkl} &= {w}_{ij}^1 {w}_{kl}^1 \nonumber
\end{align}
so that we can define a matrix $B$ that is symmetric in terms of the weights $ {w}^{sym}_{ijkl}$, ${w}^{00}_{ijkl}$ and $ {w}^{11}_{ijkl}$ so that $ \mathbf{x}^TB_{ijkl}\mathbf{x}$ is a quadratic form. The hypersurface in Eq. (\ref{eq:hypersurface}) is thus a linear combination of quadratic forms, which on itself gives a quadratic form. Depending on the data, the geometry of the separation boundary is that of circles, ellipses, lines or hyperbolas, e.g. quadric surfaces \cite{Zwillinger2002}. For the probability curves we solve $m^z=\epsilon$ analogous to Eq. (\ref{eq:mz_epsilon})
\begin{align}
    & \frac{2(h_{00}^2 + h_{10}^2)}{N} - 1 = \epsilon
\end{align}
which gives
\begin{equation}
    (h_{00}^2 + h_{10}^2) - \delta(h_{01}^2 + h_{11}) = 0
\end{equation}
with $\delta = (1+\epsilon)/(1-\epsilon)$. So the curves of equal probability are given by a linear combination of quadric surfaces.\newline

\begin{figure*}
    \centering
    \subfloat[]{
      \includegraphics[width=85mm]{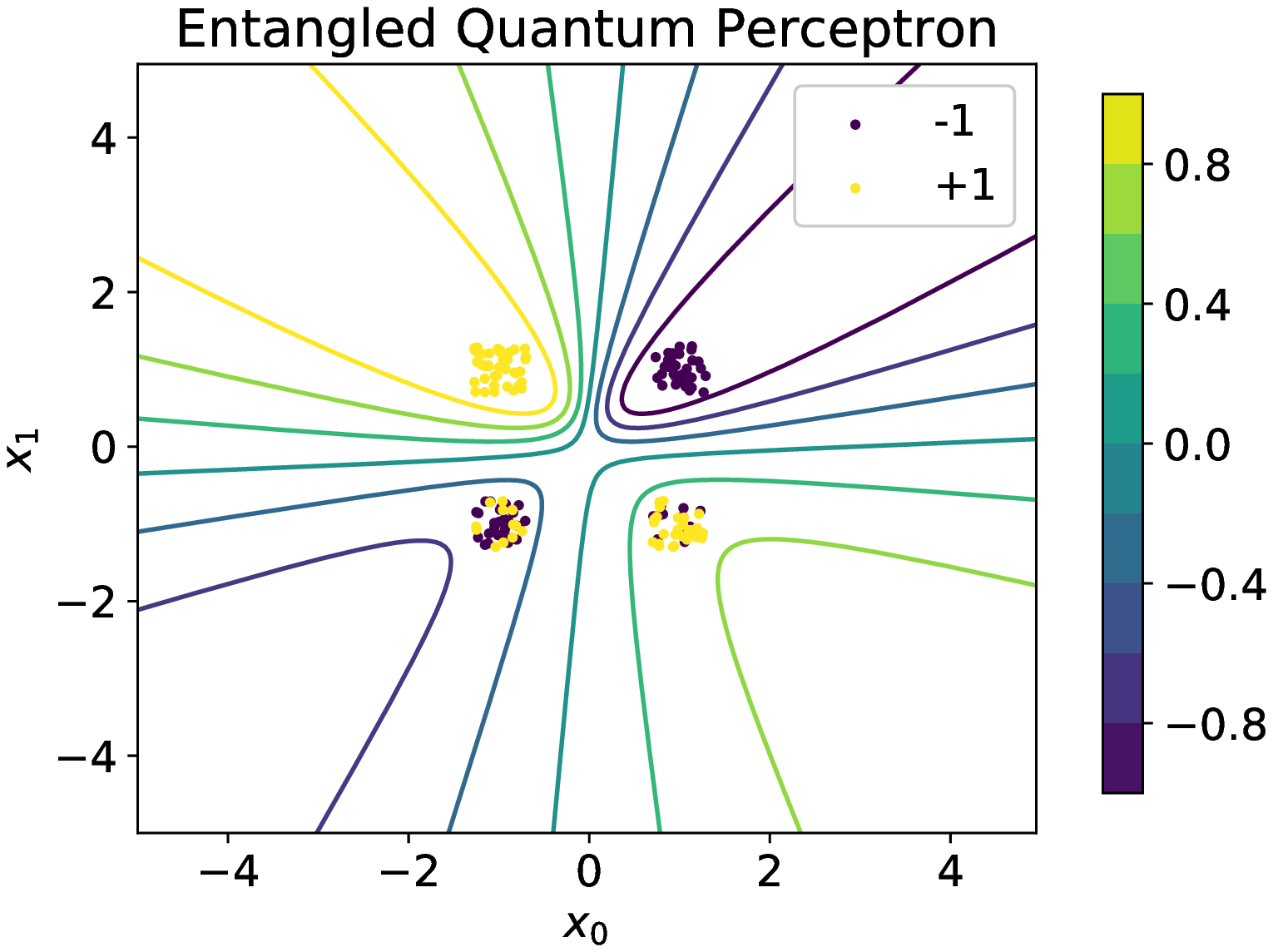}
    }
    \subfloat[]{
      \includegraphics[width=85mm]{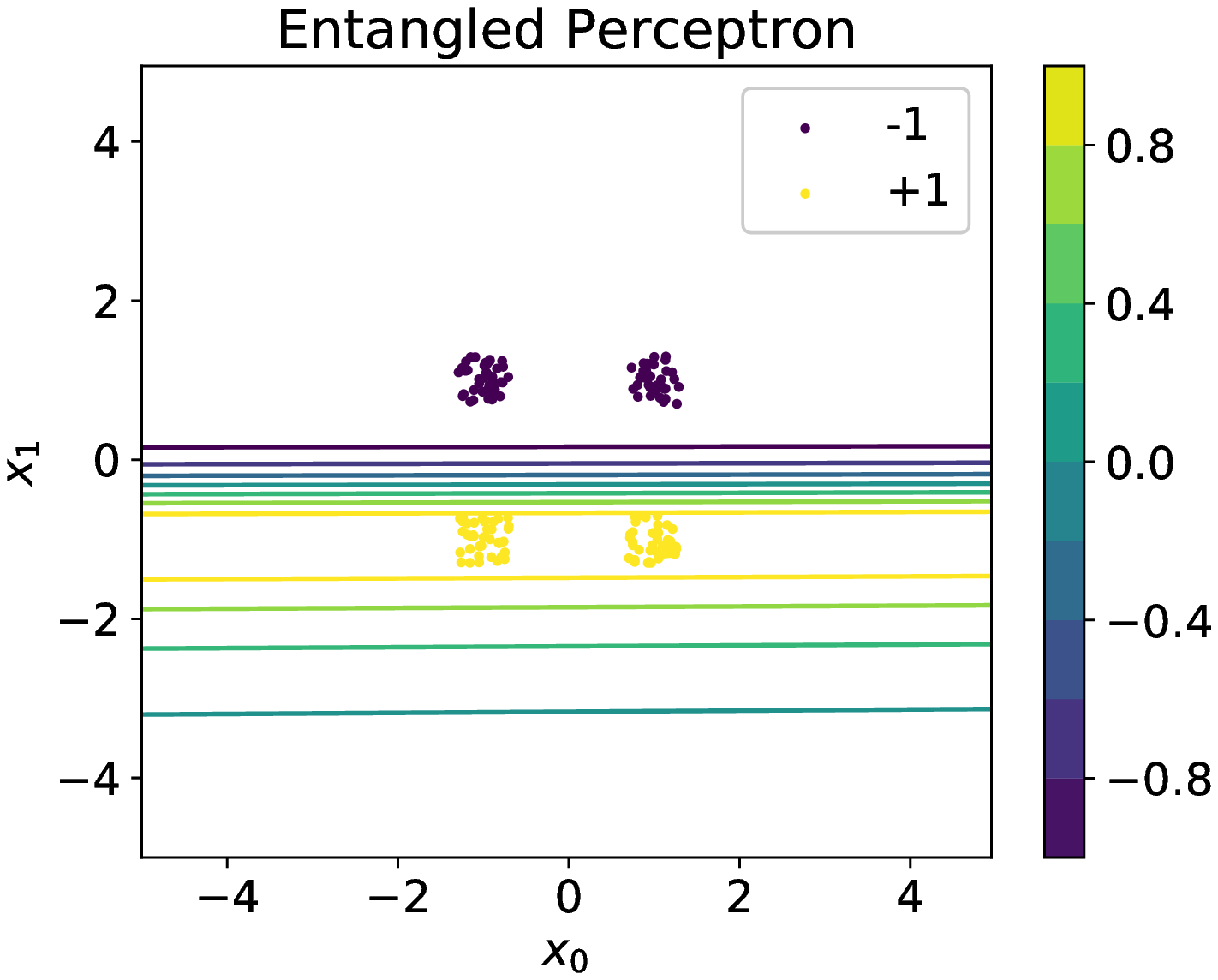}
    }
    
    \subfloat[]{
      \includegraphics[width=85mm]{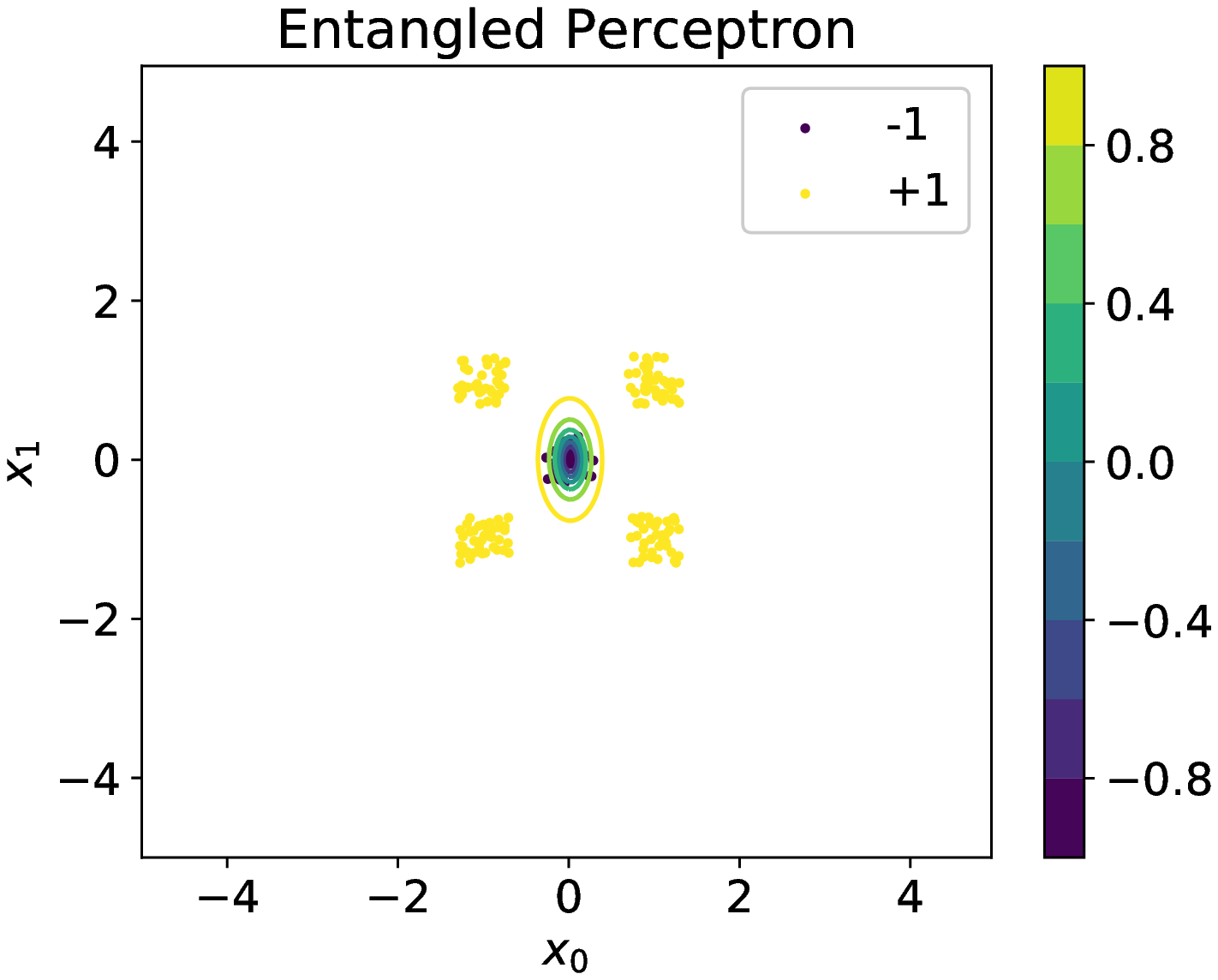}
    }
    \subfloat[]{
      \includegraphics[width=85mm]{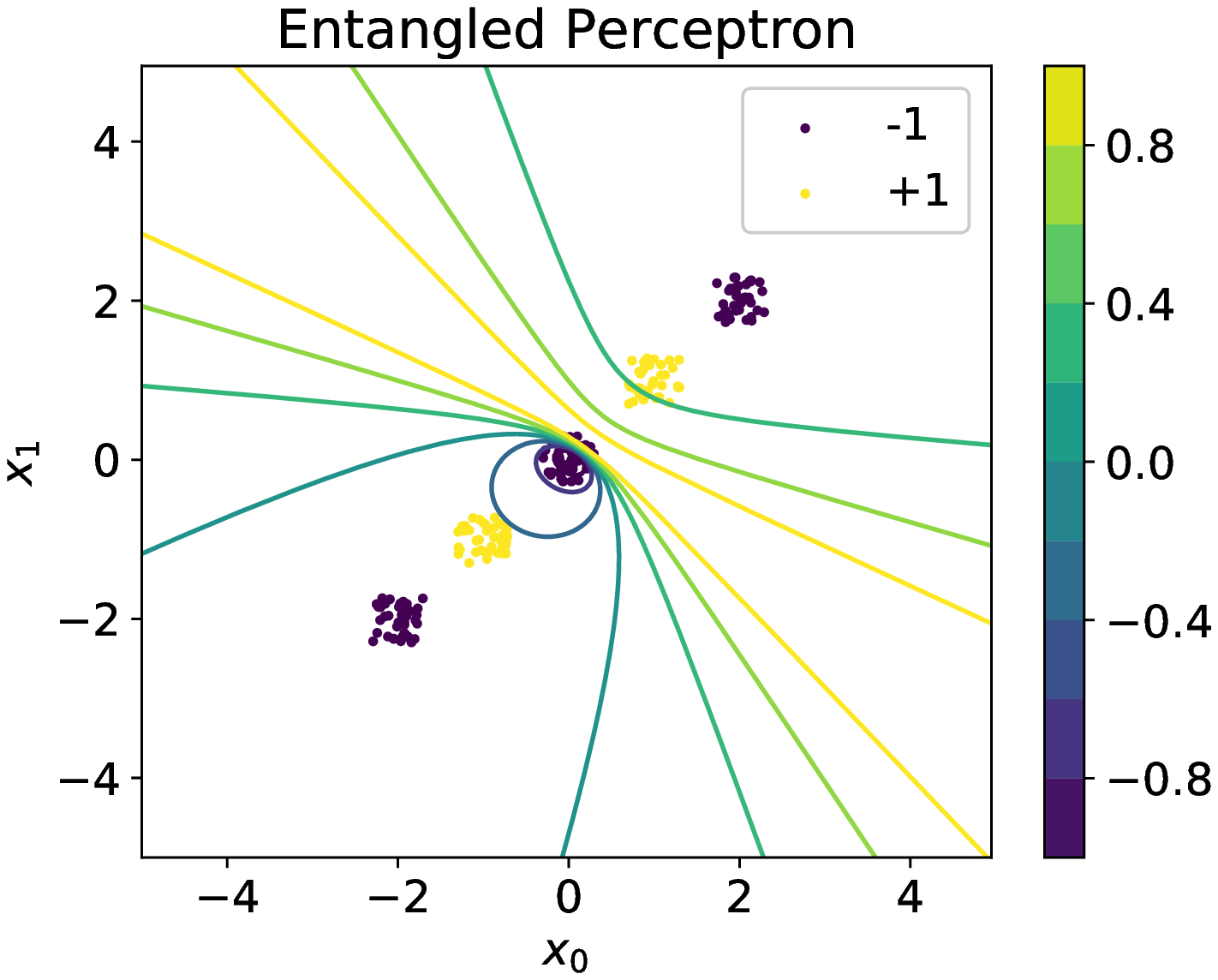}
    }
    \caption{Separation boundaries of the entangled perceptron for additional two dimensional problems. \textbf{(a)} We can still learn a noisy problem, only now we obtain quadric surfaces. This is the same problem as in Fig. 3 in the main text, except that the two bottom samples contain $30\%$ flipped labels. The two top samples are again classified perfectly, but the bottom samples get a lower probability attributed to them. \textbf{(b)} A quadric surface can also consist of parallel lines, allowing us to learn linearly separable problems. \textbf{(c)} For a setup where class $y=-1$ is surrounded by class $y=1$, we can learn an elliptical separation boundary to perfectly classify the data. \textbf{(d)} Problems that cannot be solved with a quadric surface are still problematic and lead to bad solutions.} 
    \label{fig:4_problems}
\end{figure*}

For additional examples of the quadric boundary surfaces, see Fig. \ref{fig:4_problems}.
\newpage 

\end{appendices}

\end{document}